\documentclass[aps,prb,twocolumn,showpacs,groupedaddress]{revtex4}
\usepackage{graphicx}%

\begin{document}

\title{Localization, Coulomb interactions and electrical heating in \\single-wall carbon nanotubes/polymer composites}

\author{J.M. Benoit }
\altaffiliation{Present address: MPI-FK, Heisenbergstrasse 1, 70569 Stuttgart, Germany}
\author{B. Corraze}
\author{O. Chauvet}
\altaffiliation{ Corresponding author, e-mail: chauvet@cnrs-imn.fr }
\affiliation{ Institut des Materiaux Jean Rouxel IMN/LPC,2 rue de la Houssiniere, 44322 NANTES, FRANCE }

\date{\today}

\begin{abstract}
Low field and high field transport properties of carbon nanotubes/polymer composites are investigated for different tube fractions. Above the percolation threshold $f_{c}\sim$0.33\%, transport is due to hopping of localized charge carriers with a localization length $\xi \sim$10-30 nm. Coulomb interactions associated with a soft gap $\Delta_{CG}\sim$2.5 meV are present at low temperature close to $f_{c}$. We argue that it originates from the Coulomb charging energy effect which is partly screened by adjacent bundles. The high field conductivity is described within an electrical heating scheme. All the results suggest that using composites close to the percolation threshold may be a way to access intrinsic properties of the nanotubes by experiments at a macroscopic scale.
\end{abstract}


\maketitle

Because of their structure, single-wall carbon nanotubes (SWNT) are ideal mesoscopic 1D systems. As such, their electronic transport properties are controlled by quantum size and charging effects. Most of these properties expected for 1D mesoscopic conductors have been observed on individual SWNT or bundles: Coulomb blockade and level quantization\cite{tans}, Luttinger liquid characteristics\cite{bockrath}, ballistic transport\cite{frank}.... On the other hand, measurements on macroscopic SWNT mat\cite{fisher,gaal,yosida} show classical transport properties. In most of the cases, the conductivity is hopping like at low temperature suggesting localized charge carriers without any contribution of Coulomb interactions. Indeed, the large number of tubes involved along the conduction path should be partly responsible of the disappearance of the mesoscopic and metallic characteristics. However the reasons for the loss of the Coulomb interaction effects are unclear.

In this paper, we propose to investigate these differences by studying the transport properties of SWNT/polymer composites thin films where we limit the number of tube-tube contacts by varying the SWNT content. In all samples, the conductivity is hopping like with a localization length which seems to be independent of the SWNT content. In contrast with other works\cite{fisher,gaal,yosida} Coulomb interactions associated with a soft Coulomb gap opening play a dominant role in the low temperature transport. We suggest that it originates from the Coulomb charging energy involved in the tube-tube transfer. In addition, large non linearities of the conductivity are observed at high electric field that we discuss in terms of electrical heating of the conduction gas. All our results indicate that close to the percolation threshold, some intrinsic properties of the tubes can be reached at a macroscopic scale. Coulomb interactions are essential in these systems. It corroborates the use of the Luttinger Liquid model for individual nanotubes. Still the mesoscopic character is lost because of the large number of tubes involved.

The SWNT are produced by arc discharge\cite{journet} with a typical purity of about 70-90 \%. Their characteristic diameter is in the range 1.3-1.5 nm and most of them are embedded in bundles with a typical size of 7-12 nm as shown by transmission electron microscopy (TEM). SWNT/PMMA composites are formed by drop casting a sonicated solution of SWNT, polymethylmetacrylate (PMMA) and toluene. Thin films are obtained with a thickness $\sim$10 $\mu$m. The SWNT content varies between a volume fraction $f$=0.1\% to $f$=8\%. Electron microscopies show that the films do not present large heterogeneities at their surface or in the bulk\cite{note1}. Transport measurements are performed within a planar four probe configuration with gold evaporated contacts $\sim$ 20 $\mu$m apart.

\begin{figure}
\includegraphics[bb=181 345 410 487, clip]{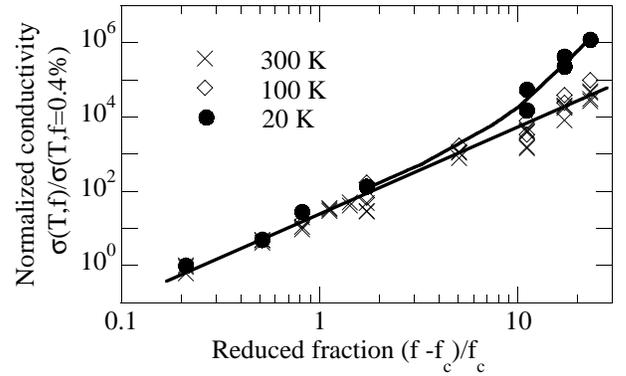}
\caption{\label{fig1} Normalized conductivity $\sigma(T,f)/\sigma(T,f=4\%)$ versus the SWNT reduced volume fraction $(f-f_{c})/f_{c}$ of the composite thin films at $T$=300 K, 100 K and 20 K respectively.}
\end{figure}
As already discussed elsewhere\cite{benoit}, the evolution of the room temperature conductivity $\sigma$ of the composite thin films with $f$ is well described within the standard percolation theory. The critical behavior $\sigma \propto (f-f_{c})^{\beta}$ with a percolation threshold $f_{c}$=0.33$\pm$0.03\% and a critical exponent $\beta$=2.1$\pm$0.1 is shown in Fig.\ref{fig1}. It is obeyed over two orders of magnitude in reduced fraction at room temperature and at 100 K. Indeed, the low value of the threshold $f_{c}$ arises from the very high aspect ratio (length over radius ratio) of the tubes which is found to be close to some hundreds\cite{benoit}. The critical exponent $\beta$=2.1 is in good agreement with the conventional 1.94 exponent found for random 3D connectivity. This quasi-ideal percolation behavior confirms that the dispersion of the conducting fillers in the matrix is good. Indeed, conduction is due to metallic SWNT\cite{sigma}. Since a percolation behavior is obeyed, we expect that close to the threshold, very few metallic tubes contribute to the conduction pathway.

\begin{figure}
\includegraphics[bb=181 303 407 534,clip]{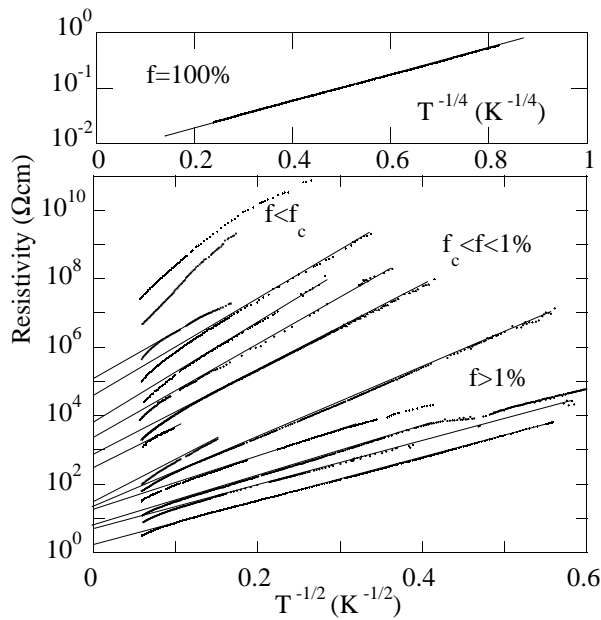}
\caption{\label{fig2}{\it Lower panel:} Temperature dependence of the resistivity versus $T^{-1/2}$ of some composite films with $f$= 0.15, 0.2, 0.35, 0.4, 0.5, 0.6, 0.9, 1, 2, 4, 4, 6, 6, 8 \% respectively from top to bottom. {\it Upper panel:} Temperature dependence of the resistivity of a SWNT pressed pellet ($f$=100\%) versus $T^{-1/4}$. The line shows the agreement with Mott's model.} 
\end{figure}
The temperature dependence of the resistivity $\rho$ of our samples is plotted in the lower panel of Fig.\ref{fig2}. The resistivity is not metallic like in the investigated temperature range. It varies with temperature $T$ according to 
\begin{eqnarray}
\rho=\rho_{0}\exp(T_{\alpha}/T)^{\alpha}
\label{eq1}
\end{eqnarray}
with $\alpha$=1/2 below 30-50 K. This is the signature of localized carriers hopping. The exponent becomes $\alpha$=1/4 at higher temperatures. The composite films can be compared to a pressed pellet {\it i.e.} an ideal $f$=100 \% sample. As shown in the upper panel, the pellet resistivity perfectly obeys a $\alpha$=1/4 law even at low temperature, in agreement with published data\cite{gaal,yosida}. The different $\alpha$ exponents as well as the slopes evolution in Fig.\ref{fig2} mean that some additional mechanisms set in at low temperature in addition to the network connectivity effect. It is reflected by the shift of the conductivity from the critical dependence shown in Fig.\ref{fig1} for $f \geq$ 1\% at $T$=20 K. It contrasts with more conventional percolating networks\cite{putten}.

Hopping transport is frequently observed in SWNT mat\cite{gaal,yosida}. In Mott's model\cite{mott} ($\alpha$=1/4), $T_{\alpha=1/4}$ is related to the localization length $\xi$ and to the density of states $n(E_{F})$ at the Fermi level. Assuming that the charge carriers originate from the metallic tubes leads to a localization length $\xi \sim$8.2 nm for the pressed pellet. Indeed, the use of Mott's model here is questionable. It requires a constant DOS at the Fermi level which is the case for metallic SWNT. It is also restricted to a single phonon process. This condition usually limits the validity of the model to low temperature. However, the Debye temperature of SWNT should be quite high\cite{debye} and thus the single phonon process is expected to extend to rather high temperatures.

\begin{figure}
\includegraphics[bb=185 353 407 498,clip]{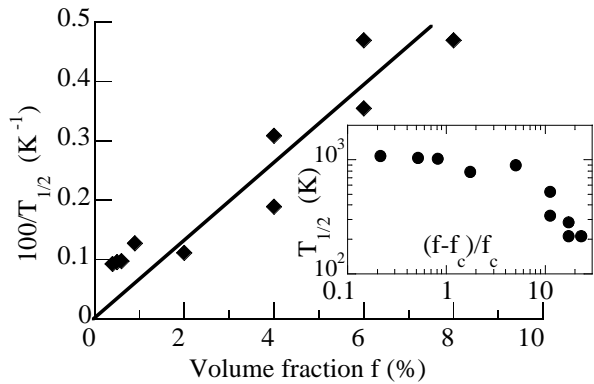}
\caption{\label{fig3}Variation of $100/T_{1/2}$ deduced from the curves of Fig.\ref{fig2} with the SWNT volume fraction $f$. Inset: $T_{1/2}$ versus $(f-f_{c})/ f_{c}$. $T_{1/2}$ is constant below $f$= 1 \%.}
\end{figure}

As already noticed, the composite films show a different temperature dependence of the resistivity, described by Eq.\ref{eq1} with $\alpha$=1/2. This behavior indicates the presence of Coulomb interactions\cite{efros} which open a soft Coulomb gap\cite{gap} $\Delta_{CG}$ at the Fermi level related to the slope of the resistivity curves $k_{B}T_{1/2}=2.8 e^{2}/\kappa\xi$ where $\kappa=\kappa_{host}+4\pi e^{2}n(E_{F})\xi^{2}$ is the effective dielectric constant ($\kappa_{host}\sim$3 here). The evolution of the experimental $T_{1/2}$ with $f$ is given in inset of Fig.\ref{fig3}. Below $f$=1\%, $T_{1/2}\sim$1000 K is almost constant. When $f$ increases up to 8\%, $T_{1/2}$ decreases down to 200 K as 1/$f$. This scaling arises from the dilution of the SWNT in the matrix where $n(E_{F})= n^{0}(E_{F}) \times f$ has to be considered instead of the pure SWNT DOS $n^{0}(E_{F})$. The scaling parameter gives a localization length $\xi \sim$7.4 nm in excellent agreement with the previous result. It indicates that the localization mechanism is intrinsic to the SWNT system independently of the $f$ fraction.

To our knowledge, while $\alpha$=1/4 hopping is widely reported in the literature on SWNT mat, this is the first time that a $\alpha$=1/2 regime is observed. It suggests that Coulomb interactions are effective in dilute SWNT systems but not in non dilute mat. Below $f$=1\% the constant $T_{1/2}$ means a constant (soft) Coulomb gap $\Delta_{CG} \simeq$2.5 meV. It decreases down to 0.5 meV when $f$ increases up to 8\%. We argue that this Coulomb gap is due to the Coulomb charging energy involved in the transport process from one bundle (or tube) to the neighboring one. The charging energy is responsible for the Coulomb blockade phenomenon reported on isolated SWNT or bundles. Its order of magnitude\cite{tans,bockrath} is 1-3 meV, in agreement with our value for $\Delta_{CG}$. When increasing $f$ in the composite, the bundles (tubes) come closer to each other and a mutual screening occurs which reduces the charging energy. A crude estimate can be given using basic electrostatic theory. We compare the charging energy $E_{c1}$ of an isolated cylindric wire (an isolated bundle) with the charging energy $E_{c2}$ of a coax (an isolated bundle surrounded by its neighbors): $E_{c1}/E_{c2}=\ln(L/R)/\ln((s+R)/R)$ where $L$ is the length of the bundle, $R$ its radius and $s$ the bundle separation. A charging energy of $E_{c1}\sim$3 meV corresponds to an isolated bundle with $R$=5 nm and $L$=2 $\mu$m. For $f$=10\%, the average separation between bundles is 18 nm assuming that they run parallel to each other. Within this simple model, a reduction of the charging energy by a factor 4 is thus expected in rough agreement with our results. Below $f$=1\%, the bundles are sufficiently far from each other for such a screening to be inefficient. Conversely increasing $f$ further leads to fully screened interactions in the case of the pellet ($f\geq$40\% gives already a screening effect by a factor of 10) and thus restores the Mott's model. These results suggest that even in macroscopic composite films, intrinsic (nanoscopic) charging effects are still detectable close to the threshold.

Still localization is observed. The fact that $\xi \sim$7-8 nm is much larger than the tube diameter (but much lower than the mean free path) suggests that localization does not occur at the tube scale. It means also that inter-tube transfers are not limiting in the investigated temperature range\cite{stahl}. Possible origins of localization are. {\it i-} The charges are localized at the bundle boundaries (transverse localization). Then $\xi$ gives an average diameter of the bundles. Our result is of the correct order of magnitude. {\it ii-} Localization occurs along the bundles by bundle-bundle (or tube-tube) contacts (longitudinal localization). Then the localization length is expected to depend on the filling of the composites. This is not observed. {\it iii-} Localization is due to defects along the tubes or into the bundles (bad matching of chirality, structural defects...). We cannot disregard this possibility. However, Raman and TEM characterizations show that the SWNT are of high quality.

\begin{figure}
\includegraphics[bb=183 352 413 494, clip]{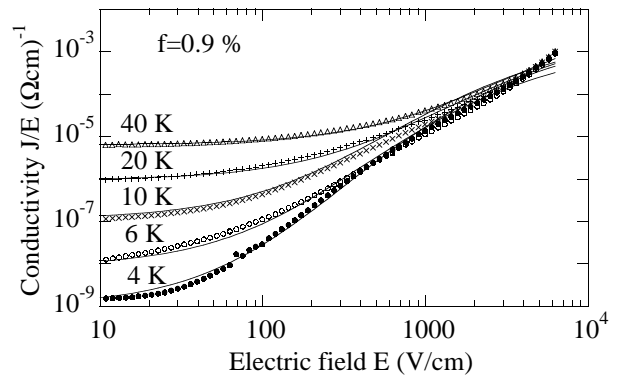}
\caption{\label{fig4} Non linearities of the conductivity $J/E$ with the electric field strength for the composite film with $f$=0.9 \% at different temperatures. The lines are the results of the fit discussed in the text. This sample is representative of the others.}
\end{figure}

Intrinsic localization is also reflected in the high field transport. Figure \ref{fig4} shows also that large non linearities are observed in the $I$($V$) characteristics of the films (shown here for $f$=0.9 \%) for moderate electric field strength $E$.
The experiments were carried out using standard pulse current techniques. Non linearities are observed in all samples\cite{f1}. The non linear effects can reach up to 6 orders of magnitude at low temperature for the smallest SWNT contents. Increasing the temperature or $f$ leads to smaller effects. Indeed, we have checked that it is not due to self-heating. In localized systems, such non linear effects can have different origins\cite{mott,jonscher}: injection of excess carriers, modification of the emission probability, modification of the conduction pathway... We will not discuss these models however they do not apply here mainly because our system consists of conducting samples with a small DOS and a large localization length. Alternatively we suggest that the non linearities are due to an energy redistribution of the carriers caused by the high electric field \cite{jonscher,shklovskii}. Such a redistribution requires that the density of states is low and that empty states exist close to the filled states which is indeed the case in metallic SWNT. It is equivalent to a thermal effect and it can be associated with an "electric temperature"\cite{shklovskii} $T_{elec}$: 
\begin{eqnarray}
k_{B}T_{elec}=\gamma eEa
\label{eq2}
\end{eqnarray}
where $e$ is the electron charge, $a$ the localization length and $\gamma$ a numerical factor close to 1. When applying high electric field at finite temperature $T$, both thermal and electrical heating play a role. The question of the effective temperature $T_{eff}$ seen by the carriers is not easy to answer\cite{cleve}. The simplest assumption is that $T_{eff}=T+ T_{elec}$. Within this model, the conduction mechanism remains the same at low or high field but the thermostat temperature $T$ has to be replaced by the effective one $T_{eff}$. The curves of Fig.\ref{fig4} have been fitted according to $J/E =\sigma_{0}\exp\{-(T_{1/2}/[T+\beta E])^{1/2}\}$. We adjust $\beta$ at each temperature $T$. As can be seen, the agreement is quite good. Unexpectedly, we find that $\beta\simeq$0.035 K.cm/V is almost constant for all the curves. It suggests that electrical heating really occurs. Typically $T_{elec}\sim$35 K when $E$=100 V/cm. Obviously at high fields, $T_{elec}\gg T$ and $J/E$ does not depend any more on $T$. According to Eq.\ref{eq2}, $\beta$ is related to a localization length. Using $\gamma$=1 gives a length $a\sim$28 nm. This value is higher but still of same order of magnitude as $\xi\sim$8 nm found from the temperature dependence of the "zero field" conductivity. It supports the use of the electric temperature concept.

\begin{figure}
\includegraphics[bb=181 336 411 510,clip]{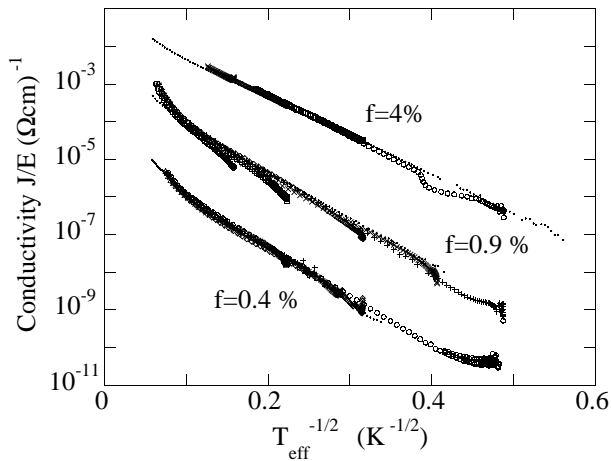}
\caption{\label{fig5} Low field $\sigma$ and high field $J/E$ conductivity versus ${T_{eff}}^{-1/2}=(T+\beta E)^{-1/2}$ of the composite films with $f$=0.4, 0.9 and 4 \% respectively. $\beta$=0.035 K.cm/V for all samples. For each set of curves, the ohmic conductivity $\sigma(T=T_{eff})$ of Fig.2 is represented by small dots while the isothermal $J/E$ are represented by large dots.}
\end{figure}

We have suggested previously that localization is intrinsic to the SWNT system and it does not depend on $f$. Following this argument, we definitively fix $\beta$=0.035 K.cm/V ({\it i.e.} the localization length) for all samples and we plot the high field conductivity $J/E$ as well as the ohmic (low field) one $\sigma$ versus ${T_{eff}}^{-1/2}$ on the same figure. The results are shown in Fig.\ref{fig5} for $f$=0.4, 0.9 and 4\% respectively.
For a given $f$ fraction, all the curves (isothermal $J(E)$ and $\sigma (T)$) merge into a {\it master plot} characteristic of the $f$ fraction. It definitively confirms that localization is intrinsic and does not depend on the way the SWNT are dispersed into the insulating matrix. It shows also that a unique transport mechanism is responsible of both the low field and the high field conductivity.

In summary, this work shows that in SWNT/PMMA composites: {\it i- }Transport is due to localized carriers originating from the metallic SWNT with a characteristic localization length of 8-28 nm due to bundle boundaries. {\it ii- }Coulomb interactions arising from the charging energy limit the transport at low temperature. These interactions are partly screened by adjacent bundles. For a sufficiently large dispersion of the SWNT, screening is ineffective and the charging energy regime found in isolated SWNT transport is recovered at low temperature. {\it iii- }The connectivity of the nanotubes network into the matrix dominates the transport properties as soon as the temperature is high enough to wash out the Coulomb gap. {\it iv- }Because of the specific DOS of metallic SWNT, high electric fields can redistribute the energy of the carriers which leads to large non linearities of the $I(V)$ characteristics. Finally this work suggests that using composites close to the percolation threshold may be a way to access intrinsic tube properties at a macroscopic level. This approach may be developed for other transport properties such as thermal conductivity.\\
We acknowledge P. Bernier for providing us with the SWNT. This work is partly supported by the EEC COMELCAN HPRN-CT-2000-00128 contract


\begin{thebibliography}
\bibitem{note1} This is not the case for larger fractions.
\bibitem{gap} If the temperature is high enough, thermal effects shadow the Coulomb gap and a classical Mott's mechanism is recovered. This is verified in our samples.
\bibitem{f1} excepted for $f$=8\% and for the pressed pellet for which self heating perturbs the measurements.
\end{thebibliography}
\end{document}